\documentclass[preprint2]{aastex}
\usepackage{graphicx}

\def\aa_{\aap}

\newlength{\bigfigwidth}
\begin{document}

\title{On the helium flash in low-mass Population~III Red Giant stars
\vspace{-3cm}\flushright{MPA 1345e}\vspace*{2.5cm}}
\shorttitle{On the He-flash in low-mass Pop.~III stars}

\author{H.~Schlattl\altaffilmark{1,2,3}, S.~Cassisi\altaffilmark{2,1},
 M.~Salaris\altaffilmark{3,1},
 A.~Weiss\altaffilmark{1,4,5}\\*[2.3cm]\centerline{submitted to: \it
 The Astrophysical Journal}\vspace*{1.1cm}}

\altaffiltext{1}{Max-Planck-Institut f\"ur Astrophysik,
Karl-Schwarzschild-Stra{\ss}e 1, D-85741
Garching, Germany}
\altaffiltext{2}{Osservatorio Astronomico di Collurania, Via M. Maggini,
I-64100 Teramo,
Italy}
\altaffiltext{3}{Astrophysics Research Institute, Liverpool John Moores
University,
Twelve Quays House, Egerton Wharf, Birkenhead CH41 1LD, UK}
\altaffiltext{4}{Institute for Advanced Study, Olden Lane, Princeton, USA}
\altaffiltext{5}{Princeton University Observatory, Peyton Hall,
Princeton, USA} 

\begin{abstract}
We investigate the evolution of initially metal-free, low-mass Red
Giant stars through the He core flash at the tip of the Red
Giant Branch. The low
entropy barrier between the helium- and hydrogen-rich layers enables
a penetration of  the helium flash driven convective zone into the
inner tail of the extinguishing H-burning
shell. As a consequence, protons are mixed into
high-temperature regions triggering a H-burning
runaway. The subsequent dredge-up of
matter processed by He and H burning enriches the stellar surface
with large amounts of helium, carbon and nitrogen.
Extending previous results by \citet{HIF90} and
\citet{FII00}, who claimed that the H-burning runaway is an intrinsic
property of extremely metal-poor low-mass stars, we found that its occurrence
depends on additional parameters like the initial composition
and the treatment of various physical processes.

We perform some comparisons between predicted surface chemical
abundances and observational measurements for extremely
metal-deficient stars. As in previous investigations, our results
disclose that although the described scenario provides a good
qualitative agreement with observations, considerable
discrepancies still remain.  They may be due to a more complex
evolutionary path of \lq{real}\rq\ stars, and/or some shortcomings in
current evolutionary models.

In addition, we analyze the evolutionary properties after the 
He core flash, during both the central and shell He-burning phases,
allowing us to deduce some interesting differences between models
whose Red Giant Branch progenitor has experienced the H-flash
and canonical
models. In particular, the Asymptotic Giant Branch 
evolution of extremely metal-deficient stars and the occurrence
of thermal pulses are strongly affected by the previous RGB evolution.
\\*[0.3cm]
\end{abstract}

\section{Introduction}

The most metal-deficient stars in our Galaxy provide astronomers with
a wealth of information about the physical conditions of the universe
soon after the Big Bang in the
early epoch of galaxy formation.
In particular, the first generation of stars ---
commonly called \lq{Population III}\rq --- which formed from
primordial, essentially metal-free matter, should provide the
oldest bright objects in 
galaxies. Therefore, the identification and the
analysis of the properties of this class of stars should
offer an unique tool for investigating the chemical and physical
conditions in the very early history of our universe.

Owing to their importance for several fundamental cosmological questions,
several attempts have already been made to find zero-metal
and/or extremely metal-poor stars. The first
extensive search for Pop.~III stars has been  performed early by
\citet*{Bond}, who discovered only a few stars with metallicity
${\rm [Fe/H]} =-3$ (i.e.\ a iron content of only 1/1000 of the solar
value); more recently the sample of
extremely metal-poor stars has significantly been increased, mainly
due to extensive Wide-Field
Objective-Prism Surveys, and especially due to the Preston-Shectman
survey \citep*[usually also called HK survey]{BPS92}. Although
none of these stars shows  a \lq{true}\rq\ Pop.~III chemical
composition, there is an increasing number of stars with metallicity
${\rm [Fe/H]} <-3$, for which accurate post-detection analysis can be
performed. 
In addition, one can expect that forthcoming high resolution
spectroscopy from the
new generation of $10\,$m-class telescopes will
lead to a big improvement in the search for
this class of peculiar objects. 
Undoubtedly, unprecedented results in this field
will be achieved with planned space-based survey instruments such as
GAIA. 

In order to fully understand the results
of these surveys, it is necessary to 
investigate theoretically  not only the
main evolutionary properties of low-mass metal-free stars,
but also the changes of the surface chemical composition appearing
during that evolution. These may
be caused by accretion of metal-rich matter from the
interstellar medium and/or by internal mixing of matter processed by
nuclear reactions.
One significant observational evidence, which has yet to be
explained, is that the number of carbon-enhanced stars increases
considerably when decreasing the metallicity below
${\rm [Fe/H]}\approx-2.5$
\citep*{RBS99}. Several of the extremely metal-deficient carbon stars
show a large carbon abundance as well as nitrogen enrichment ([C/Fe]
resp.~${\rm [N/Fe]}\ga2$). In some of these
carbon-enriched stars the abundances of $s-$process elements are
enhanced, too (see \citealt*{H00} and reference therein).
This obviously implies that in extremely metal-poor stars 
mechanisms are at work which increase 
the probability for producing carbon-rich objects.

In the last decades, several theoretical
studies of the structural and evolutionary properties of
extremely metal-poor and metal-free stars (see for
instance
the reviews by \citealt{VC} and \citealt{CC} and references
therein) have shown how the
lack of heavy elements --- mainly of CNO-cycle catalysts --- produces
evolutionary features, which are rather peculiar in comparison to
stars of finite initial metallicity. One of these properties --
investigated by Fujimoto and coworkers \citep*{FIH90,HIF90,FII00} -- is
the evolution during the He-flash. Like their more metal-rich counterparts,
metal-free stars with a mass of about $1.0\,M_\odot$ reach the
conditions necessary for igniting  
triple-alpha reactions inside the He core at the tip of the
red giant branch(RGB). The electron degeneracy of the
surrounding material leads to a runaway of the He burning, the 
He core flash.
The quoted authors discovered that the small entropy barrier
between hydrogen- and helium-rich material in Pop.~III stars allows the
convective zone, which is produced by the huge energy release of
He burning, to
penetrate the overlying hydrogen-rich layers. The resulting inward
migration of 
protons into high-temperature regions (referred to as \lq{He-Flash
Mixing}\rq,
hereinafter HEFM) leads to a H shell flash.
As a consequence, when the convective envelope is deepening and
merging with the H-flash driven convective zone (HCZ) later on, 
the surface is enriched with a large amount of matter that
has experienced He-burning reactions and furthermore has been 
processed in H fusion during the H-flash. Thus,
the chemical composition of the initially metal-free stellar surface
is strongly altered and in particular
carbon-enriched.
These results provide an attractive working hypothesis for
explaining the peculiar
chemical patterns observed in extremely metal-deficient stars.

In a previous paper \citep[hereinafter Paper I]{WCSS00}, we have
investigated the evolutionary properties of low-mass metal-free
objects from the Zero Age Main Sequence (ZAMS) to the He-flash at
the tip of the RGB, using updated physical inputs and, for the first
time, accounting also for atomic diffusion. 
In view of the peculiar structural properties which
characterize Pop.~III stars \citep*[see][Paper I]{CC93},
special care was taken to use an appropriate and complete nuclear
reaction network.
Furthermore, numerical experiments were performed with the aim of
testing how the evolution of a metal-free star is affected 
when its surface is polluted by metal-rich matter
through encounters with interstellar gas clouds --- a scenario
suggested in order to explain the absence of truly
zero-metal stars and the relative paucity of extremely metal-poor
objects \citep*{Y81}.
We found that the metals in the accreted matter
fail to reach the nuclear-burning regions, so that the main structural and
evolutionary properties of the stars are not affected by the
external pollution; it was also clearly 
shown that it is quite difficult to
discriminate observationally between a polluted Pop.~III star and a very
metal-poor Pop.~II object in this phase of evolution.

In Paper I we stopped the computations soon
after the onset of the He-flash and did not
investigate the evolution during the further development of the He
flash at the RGB tip. This had been done by, e.g., \citet{FIH90}, but
until now these results have not been supported by
independent investigations\footnote{In a different context, results
similar to the one discussed by Iben and coworkers has been obtained
by \citet*{CCT}; here the He shell flash in an $0.8\,M_\odot$ model
during the thermal-pulse phase on the Asymptotic Giant Branch was
discussed.}.  
In the present work we present a reanalysis of the evolution during the
He-flash phase, covering a larger parameter space with respect to the
work by Fujimoto and coworkers. 

The main characteristics of the evolutionary code
used in the present investigation are discussed in the next section;
special emphasis is laid upon the numerical algorithms adopted to follow
the evolution during the He-flash, and the most relevant differences
with respect to previous investigations are indicated.
In \S~\ref{HeMix} the evolution through the He core flash in some
selected models is presented together with a discussion of the dependence
of the subsequent evolution on the \lq{initial conditions}\rq, like 
stellar mass and chemical composition. In the same section, we
discuss the change in the surface chemical abundances caused
by the appearance of a H shell flash like the one investigated by
\citet{FIH90}.
In \S~\ref{PostHe} we extend the computations for some
selected models, which experience a H shell flash, to the following
evolutionary phases; we discuss the evolution all
along the He-core and -shell burning phase, and compare our
results with those corresponding to He-burning models in case that
no H shell flash in the HEFM had occurred. A first comparison with
observed extremely metal-poor objects is made in \S~\ref{cobs}.
Conclusions and a brief discussion close the paper.

\section{The stellar evolution code.}

All calculations reported about in this paper were
done with the Garching stellar evolution code \citep*{wsch:2000}, which is based on the
original program developed 
by~\citet*{KWH}. The program in its present version is capable of
calculating precise solar 
models~\citep*{S00,SW99} as well as following the evolution of
low-mass stars into the latest phases of their
evolution~\citep*{JPhD,SW99}. 

\subsection{Numerical details.}\label{numtreat}
The evolution
through the He-flash can be followed in an acceptable amount of
computing time thanks to 
the grid routine implemented by \citet*{WaWe}, which ensures a
sufficiently high accuracy in the linearization of the stellar structure
equations.
Moreover, the algorithm to determine the
under-correction factor, i.e.~the actually applied fraction of the
correction found in each iteration step, 
has been improved considerably. By further limiting the time-step with
a constraint on the change of the He-burning luminosity ($L_{\rm
He}$), the critical phase of the
He-flash ($\log(L_{\rm He}/L_\odot) >7$) can be followed with about
200 models, and with at most about 200 iterations per model around the
maximum He luminosity ($\log(L_{\rm He}/L_\odot)\approx 10$).

The helium flash induced mixing found by \citet{FII00} results in a
qualitative agreement with the observed surface chemical patterns of
stars in the 
HK survey; however, their theoretical
investigation has been performed by using a limited reaction network
and a crude procedure to follow the fast mixing process during the
HEFM episode. On the other hand the
development and the general properties of the HEFM
phenomenon strongly depend on where, inside the structure, the
ingested hydrogen is burnt. \citet{FII00} assumed that the hydrogen engulfed
by the convective zone is mixed instantaneously
down to a position
at which the lifetime of a proton against capture in the
$^{12}\mathrm{C}(p,\gamma)^{13}\mathrm{N}$ reaction is equal to
the timescale of 
convective mixing. We wish to remark that in a 
previous paper, \citet{HIF90} stated that they adopted a time-dependent
mixing scheme which treats convection as a random-walk processes with
the rate of convective cell diffusion being a function of the local
mixing length and of the local buoyancy velocity. It is not clear
to us whether \citet{FII00} still use this approach, or whether they
assume instantaneous mixing.

In order to determine how deeply
hydrogen is mixed during the HEFM into the helium flash driven
convective zone (HECZ), it is crucial to treat
the mixing as a time-dependent process and to consider carefully the 
simultaneous mixing and burning processes.
In the present investigation, we therefore improved
the treatment of mixing and burning in our program.
The equations for microscopic diffusion, convection (see below) and
nuclear burning 
are solved in one common scheme \citep{PhD}; this also
implies that the hydrogen and helium networks are solved
simultaneously. 
This is necessary for the correct treatment of the nuclear evolution
during those phases, when 
H and He burning are operating at the same time, as is the case in the
core of evolved low-mass 
main sequence-stars \citep[``CNO-flash'',][]{WCSS00,FIH90} of Pop.~III
and during the HEFM.  

In our approach convection is treated as a fast
diffusive process where the diffusion constant is proportional to the
convective velocity obtained from a convection theory~\citep*{Lan85}.
When the HECZ touches the H-rich
layers at the lower boundary of the H-burning shell, the
ingested hydrogen interacts with $^{12}$C to produce $^{14}$N (CNO\,I),
releasing a huge amount of energy ($\log (L_{\rm H}/L_\odot)
\ga 6$). The movement of the outer boundary of the convective zone
is not smooth, but proceeds discontinuously and
therefore this phase is difficult to follow numerically.
Even if the number of grid points is increased considerably, the
step-wise motion could not be suppressed totally.
Consequently it is very difficult to determine in
advance when the next H-rich shell is mixed into the hotter regions
and the appropriate next time-step is nearly impossible to determine.
Using time-steps too large shortly before the onset of mixing,
the violent H-burning phase may be missed, while time-steps too
short make the 
computation of the whole phase very time consuming.

The reason for this behaviour is that
the stellar evolution program calculates the physical structure
of a model at a certain epoch and afterwards uses this structure (in
particular the density- and temperature-profile) to calculate the
chemical evolution during a certain time interval following. The solution
of all the stellar structure equations (physical as well as chemical
variables) in a common scheme would be possible, but it is
not advisable, since during this phase already a large number of
iterations ($\ga 40$) are
needed for the physical variables alone. Including the chemical evolution in
the solution of the stellar structure
would increase the computing time by a factor of
three\footnote{4 physical variables ($L$, $P$, $T$, $r$) plus 9
chemical species ($^1$H, $^3$He, $^4$He,  $^{12}$C, $^{13}$C, $^{14}$N,
$^{15}$N, $^{16}$O, $^{17}$O)}.

The separation of the chemical and structural part, dictated by
computational efficiency, however has the consequence
that, during the solution of the physical structure, any additional
mixing which is related to the movement of the boundary of the
convective zone is not taken into account. Hence, when the upper
boundary of the HECZ moves outward, the program treats 
hydrogen burning for the current position only, yielding smaller
energies than when additional hydrogen is mixed down into hotter
regions from layers just becoming convectively unstable.

If the time step during this phase is chosen too large, all protons will
be captured completely in the computation of the chemical evolution. 
The energy released, however, will be that of the total (and larger) time
step. The specific energy release per second is therefore lower than
it actually would be if the additional mixing due to the advance of
the convective boundary had been included.
With very short time-steps one could better follow the movement
of the convective zone and thus changes in the hydrogen luminosity.
However, experience shows that the (numerically) still
discontinuous penetration of the HECZ into
H-rich layers leads to oscillations in the hydrogen luminosities,
and the resolution of these would demand a disproportionately high
number of time steps.

Whenever the He-flash driven convective zone penetrates into a region
of non-vanishing hydrogen abundance, additional energy is released on
a short time scale. As long as the hydrogen abundance is smaller than
approximately 0.05, the additional energy from the violent hydrogen
fusion is negligible, because of the fact that the helium luminosity
is the dominant energy source. When the HECZ is advancing to regions
with hydrogen abundances higher than about 0.05, the large
H abundance in this zone causes a significant increase of the
H-burning rate, 
which drives a further expansion of the convective zone (see
Fig.~\ref{fig4}). The numerical code then has
to follow the ensuing H-burning runaway ($\log (L_{\rm H}/L_\odot)
\ga 10$), during which the subsequent expansion of the star may even
lead to a cessation of helium burning.

We found that the most appropriate quantity to determine the onset of
the H-flash is 
\begin{equation}\label{deltaL}
\delta L = \langle L_{\rm nuc}(\Delta t)\rangle/L_{\rm nuc}(t_0),
\end{equation}
where $L_{\rm nuc}(t_0)$ is the nuclear luminosity as calculated in the
solution of the stellar structure equations, and $\langle L_{\rm
nuc}(\Delta
t)\rangle$ is the average nuclear luminosity derived from the chemical
changes within $\Delta t=t_1-t_0$.
This quantity effectively controls the consistency between the
instantaneous energy generation and the subsequent chemical evolution
based on it. 
Whenever $|1-\delta L |$ exceeds a certain limit (presently set to about 0.7),
the chosen time-step is immediately reduced and the chemical evolution
is recalculated. The time-step is decreased until $|1-\delta L | $ becomes
sufficiently small.
With this and additional criteria for the determination of time steps
and grid resolution, it is
possible to follow numerically the He-flash and HEFM evolution of
low-mass 
metal-free stars both accurately and efficiently.

\begin{figure*}[t]
\epsscale{2.1}
\plotone{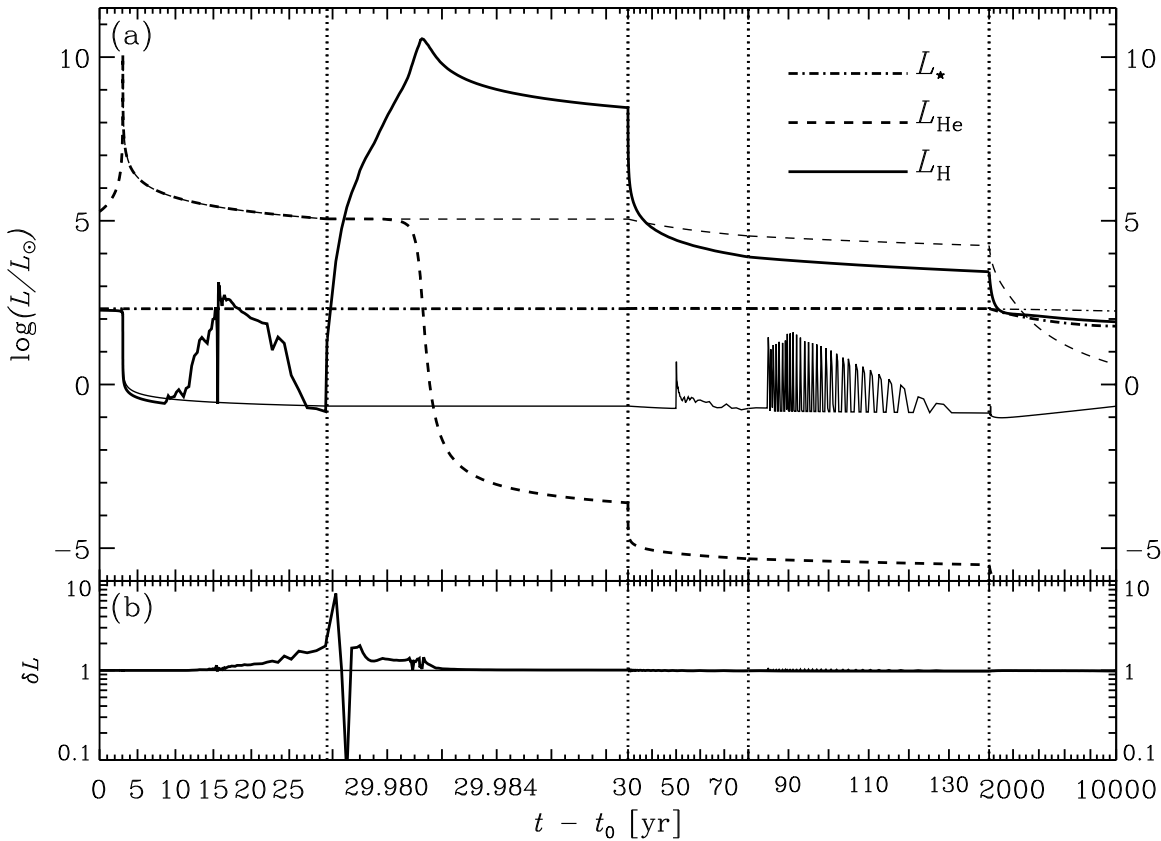}
\caption{(a) The evolution of the 
total, the H-burning, and the He-burning luminosities during the major
He core flash 
in our standard model experiencing the H-flash (thick lines, $t_0 =
6.9\,{\rm Gyr}$), and in the model without H-flash 
(thin lines, $t_0=6.5 \,{\rm Gyr}$).
(b) $\delta L$ (Eq.~\ref{deltaL}), the ratio between the mean nuclear energy
release during each time-step and the initial one. This quantity is
used to detect the onset of the H-flash in the 
computations (see text for more details). Note the changes in the x-axis scale
which are delimited by the vertical dotted lines.\label{fig1}}
\end{figure*}

In the lower panel of Fig.~\ref{fig1}  $\delta L$ during
the approach of the HECZ to the H-rich
layers is shown. Already at $t-t_0 \approx 10\,{\rm yr}$, i.e.\ before
$L_{\rm H}$ exceeds $L_{\rm He}$, hydrogen is mixed into the
He-burning convective zone, causing an increase in $\log (L_{\rm
H}/L_\odot)$ from initially about -0.5 to 2.5 (thick solid line in
upper panel of Fig.~\ref{fig1}). This energy release is, however,
negligible compared to the energy produced by the fusion of
helium. Hence, even
with relatively large time-steps, $|1-\delta L|$ remains below 0.5.
When the amount of ingested hydrogen is sufficient to
affect the subsequent evolution ($t-t_0
\approx29 \,{\rm yr}$), $\delta L$ becomes larger than	
2, leading to a drastic decrease of the time-step. The subsequent 
evolution of $L_{\rm H}$ and $L_{\rm He}$ is resolved accurately
($|1-\delta L|<0.5$).

\subsection{Input physics.}\label{inphy}
In order to describe the chemical evolution of the star we 
follow the abundances of
$\mathrm{H}$, $^3$He, $^4$He, $^{12}$C, $^{13}$C, $^{14}$N, $^{15}$N,
$^{16}$O, 
$^{17}$O, without assuming  {\it a priori} equilibrium
compositions for these chemical elements.

Since the metal distribution in our
models can show strong deviations with respect to the usually adopted
scaled-solar mixture, we decided not to use the most updated
equation of state \citep*[like OPAL-EOS;][]{OPEOS}, but the
Saha-equation for the outer stellar layers, and a simplified equation
of state for a degenerate electron gas in
the core regions~\citep{KW}\footnote{The dependence of the results
on this choice will be discussed in the next section.}.
The switching point between these two equations
of state is determined by the position where either carbon or helium
(in case there is no carbon present) is fully
or maximally ionized. 

For the radiative opacity we have used,
unless stated differently (see below), the tables provided
by \citet*{OP96} for temperatures above $T>10000$~K, and
for low temperatures ($T<8000$~K) those
by \citet*{Alex}. In the intermediate temperature range a smooth
transition between the two tables is used. Both opacity table sets
were computed by using a 
scaled-solar mixture for the heavy-element distribution. For
electron conduction opacities we used a program employing the results
by \citet{imii:83}, which allows the calculation of the opacity for
arbitrary abundances of the most important elements.

The diffusion constants for the gravitational settling of heavy elements
are calculated by a routine provided by Thoul \citetext{private
communication}, which solves Burgers' equation for a multicomponent
fluid~\citep*{diffc}. For the treatment of convection 
we apply the mixing-length
theory~\citep*{MWT} with the parameter $\alpha$=1.6. The energy losses
from photo-, pair- and plasma-neutrinos are calculated according to
\citet*{MKI85}. Mass loss has been accounted for by using the Reimers'
formula \citep*{reimers} with $\eta$=0.4.

\section{The helium flash induced mixing.}\label{HeMix}

The evolutionary and structural properties of low-mass,
metal-deficient stars have been described at several instances (see
for instance \citealt{CC93} and Paper I) and will not be
repeated here. We wish to focus our attention on the development
of the He-flash induced mixing episode and, primarily, on its dependence
on stellar parameters such as the initial helium abundance.
The occurrence and the details of the HEFM phenomena have been
discussed in detail by~\citet{HIF90}. Here we
emphasize the differences between their results and ours.

\subsection{The hydrogen flash.}\label{mainflash}

\begin{figure}[t]
\includegraphics[width=\columnwidth]{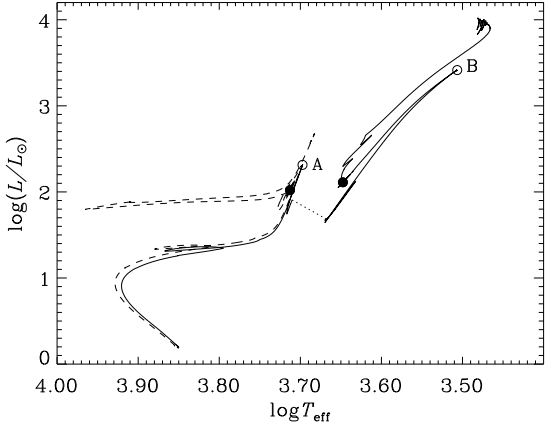}
\caption{The evolution in the H-R diagram of a $1\,M_\odot$
star with 
initial helium abundance $Y_i=0.23$ (solid line) and $Z=0$ from the
ZAMS up to the helium shell-burning phase. 
The open circles labelled by A and B correspond to the
onset of the first (major) and the second He-flash, respectively. 
The fast transition from A to a phase of slow shell burning is
indicated by the dotted segment. Also
shown, for comparison, is the track of the same model, but  
with an initial helium abundance of 0.24 (model S3), which does not
experience a hydrogen flash. The filled circles
indicate the beginning of the central
He-burning phase.\label{fig2}}
\end{figure}

Our \lq{standard}\rq\ model is a metal-free one
with mass equal to $1\,M_\odot$ and an initial helium abundance
$Y_i=0.23$, computed by neglecting both atomic diffusion and
possible external pollution. The main properties of this
model are listed in the first line of Table~\ref{tab1}. In this table
we report initial conditions,
main assumptions which characterize each case, the
information whether the HEFM leads to a H-flash, and the final
relevant surface abundances at the end of the H resp.~He-flash
for all the experiments performed (\S~\ref{Hflash}). 
The evolution of our standard model in the H-R diagram is shown in
Fig.~\ref{fig2}.


\begin{deluxetable}{lllccccccrrllc}
\tablecolumns{14}
\tabletypesize{\scriptsize}
\tablewidth{0pt}
\tablecaption{Main properties of various numerical
experiments.\label{tab1}}
\tablehead{
\colhead{} &  \colhead{} & \colhead{} & \multicolumn{5}{c}{Physics}  &
\colhead{} &
\multicolumn{5}{c}{Properties\tablenotemark{a}} \\
\cline{4-8} \cline{10-14} \\
\colhead{Mod.\tablenotemark{b}}  & \colhead{$Y_i$} &
\colhead{$\case{M_\star}{M_\odot}$} &
\colhead{$\case{D_C}{D_{\rm MLT}}$\tablenotemark{c}} &
\colhead{$\case{D_{\rm mic}}{D_{\rm Th}}$\tablenotemark{d}} &
\colhead{Poll.\tablenotemark{e}} &
\colhead{EOS} & \colhead{$\kappa$\tablenotemark{f}} &&  \colhead{$L^{\rm
max\,}_{\rm He}$\tablenotemark{g}} &
\colhead{$Y^{S\,}$\tablenotemark{h}}& \colhead{$X^S_{\rm C}$}  &
\colhead{$X^S_{\rm N}$} & \colhead{$\case{X^S(Fe)}{X^S_i(Fe)}$}
}
\startdata
S1$^\dagger$  & 0.23 & 1.0 &   1      &  ---  &  no   &  Saha &  solar
&& 10.11 & 0.477 & 0.0088   & 0.0043  &  \nodata    \\
S2$^\dagger$ & 0.235 & 1.0 &   1      &  ---  &  no   &  Saha &  solar
&& 10.06 & 0.478 & 0.0085   & 0.0040  &  \nodata    \\
S3           & 0.24  & 1.0 &   1      &  ---  &  no   &  Saha &  solar
&& 10.02 & 0.241 & 0.0      & 0.0     &  \nodata    \\
S4           & 0.25  & 1.0 &   1      &  ---  &  no   &  Saha &  solar
&& 9.92  & 0.251 & 0.0      & 0.0     &  \nodata    \\
C1$^\dagger$ & 0.23  & 1.0 &$1^{\rm CM}$&---  &  no   &  Saha &  solar
&& 10.10 & 0.472 & 0.0086   & 0.0041  &  \nodata    \\
O1$^\dagger$ & 0.23  & 1.0 &   1      &  ---  &  no   &  Saha &  C\&N
&& 10.10 & 0.466 & 0.0088   & 0.0035  &  \nodata    \\
E1$^\dagger$ & 0.23  & 1.0 &   1      &  ---  &  no   &  OPAL &  solar
&& 10.54 & 0.436 & 0.0080   & 0.0032  &  \nodata    \\
A1$^\dagger$ & 0.23  & 1.0 &   0.01   &  ---  &  no   &  Saha &  solar
&& 10.10 & 0.399 & 0.0060   & 0.0026  &  \nodata    \\
A3           & 0.24  & 1.0 & $10^{3}$ &  ---  &  no   &  Saha &  solar
&& 10.02 & 0.241 & 0.0      & 0.0     &  \nodata    \\
D1$^\dagger$ & 0.23  & 1.0 &   1      &  1.0  &  no   &  Saha &  solar
&& 10.13 & 0.469 & 0.0064   & 0.0069  &  \nodata    \\
D3$^\dagger$ & 0.24  & 1.0 &   1      &  1.0  &  no   &  Saha &  solar
&& 10.05 & 0.471 & 0.0066   & 0.0066  &  \nodata    \\
D4           & 0.25  & 1.0 &   1      &  1.0  &  no   &  Saha &  solar
&& 9.96  & 0.240 & 0.0      & 0.0     &  \nodata    \\
D+4$^\dagger$& 0.25  & 1.0 &   1      &  5.0  &  no   &  Saha &  solar
&& 10.09 & 0.459 & 0.0092   & 0.0035  &  \nodata    \\
P1           & 0.23  & 1.0 &   1      &  1.0  &  yes  &  Saha &  solar
&& 10.04 & 0.232 & $1.4\times10^{-4}$ & $3.4\times10^{-5}$ &  0.032  \\
P3           & 0.24  & 1.0 &   1      &  1.0  &  yes  &  Saha &  solar
&& 9.96  & 0.241 & $1.4\times10^{-4}$ & $3.4\times10^{-5}$ &  0.032  \\
DP1$^\dagger$& 0.23  & 1.0 &   1      &  1.0  &  yes  &  Saha &  solar
&& 10.09 & 0.464 & 0.0086   & 0.0040  &  0.013    \\
DP3          & 0.24  & 1.0 &   1      &  1.0  &  yes  &  Saha &  solar
&& 10.0\phn  & 0.230 & $1.4\times10^{-4}$ & $3.5\times10^{-5}$ &  0.032  \\
M1$^\dagger$ & 0.23  & 0.82&   1      &  ---  &  no   &  Saha &  solar
&& 10.30 & 0.512 & 0.0069   & 0.0098  &  \nodata    \\
M3$^\dagger$ & 0.24  & 0.81&   1      &  ---  &  no   &  Saha &  solar
&& 10.25 & 0.550 & 0.0086   & 0.0097  &  \nodata    \\
B1$^\dagger$ & 0.23  & 0.82&   1      &  1.0  &  yes\tablenotemark{i}
&  Saha &  C\&N &&   10.34 & 0.520 & 0.0123 & 0.0050 & $3.4\times10^{-4}$
\\
B2$^\dagger$\tablenotemark{j} & 0.23  & 0.82&   1     &  1.0  &
yes\tablenotemark{i}
&  Saha &  C\&N &&   10.31 & 0.502 & 0.0115 & 0.0048 & $3.9\times10^{-4}$
\enddata
\tablenotetext{a}{Abundances after dredge-up (H-flash) resp.~at beginning
of HB}
\tablenotetext{b}{Models experiencing a H-flash are denoted by $^\dagger$.}
\tablenotetext{c}{Ratio of actual diffusion constant to the one
derived from mixing length theory \protect\citep[$\alpha_{\rm
MLT}=1.6$]{MWT}. ``CM''
marks model calculated with the theory
of~\protect\citet[$\alpha_{\rm CM}=0.8$]{CM1,CM2}.}
\tablenotetext{d}{Ratio of actually applied microscopic diffusion
constant to the values derived by \protect\citet{diffc}.}
\tablenotetext{e}{$Z=0.02$, $\Delta M=0.01\,M_\odot$ (mass of
additional, polluted matter)}
\tablenotetext{f}{Heavy element distribution used in the opacities:
\lq{solar}\rq\ and \lq{C\&N}\rq\  mean that a scaled-solar
resp.~carbon- and nitrogen-enhanced mixture has been adopted.} 
\tablenotetext{g}{Logarithm of the maximum He-burning luminosity in
units of the solar luminosity.}
\tablenotetext{h}{S=surface, i=initial}
\tablenotetext{i}{In this sequence, the value of $\Delta M=3\times
10^{-4}\,M_\odot$ has been chosen in order to match the observed [Fe/H]
value in the star CS22892-052 (see text for more details).}
\tablenotetext{j}{No mass-loss in this model, i.e.~$\eta_{\rm
Reimers}=0$}
\end{deluxetable}

\begin{figure}[b]
\includegraphics[width=\columnwidth]{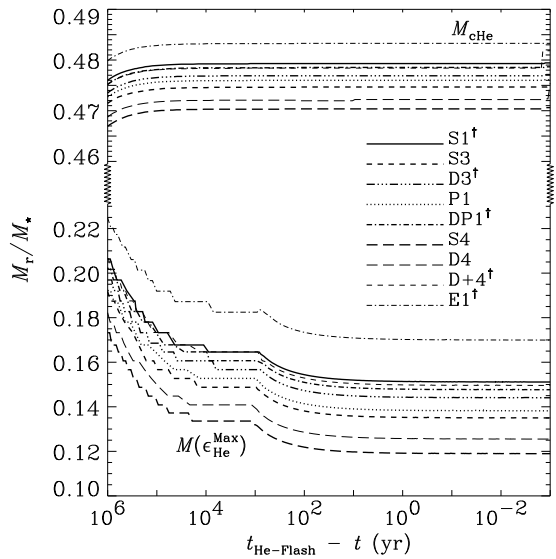}
\caption{The evolution of the He core mass ($M_{\rm cHe}$) and
of the mass shell where
the He-burning energy release is at maximum [$M(\epsilon^{\rm He}_{\rm
max})$] for different model sequences.
The notation of the models follows
Table~\ref{tab1}, where the input physics of the various sequences is
defined. The solid line corresponds to
our standard model. 
The symbol $\dagger$ is used to identify the models
experiencing the H-flash. 
\label{fig3}}
\end{figure}

The major He core flash occurs when the surface luminosity is
equal to $\log(L/L_\odot)=2.314$, i.e.~a factor of $\approx 1.5$ lower than
the value attained by the model of Iben and coworkers. At this stage,
the mass of the He core is $M_{\rm cHe} = 0.482\,M_\odot$. The
mass shell of maximum energy release by He burning,
$M(\epsilon^{\rm He}_{\rm max})$, is located at
$M_r=0.151\,M_\odot$. These values are in fair agreement with the ones
obtained by~\citet{CC93}. The evolution of $M_{\rm cHe}$ and
$M(\epsilon^{\rm He}_{\rm max})$ is shown in Fig.~\ref{fig3}.

Significantly different results were found by
~\citet{FIH90}. In their model (of same mass) the size of the H-exhausted core at
He-flash ignition is $0.528\,M_\odot$ and
$M(\epsilon^{\rm He}_{\rm max})$ is $0.41\,M_\odot$.
We do not know the reason for such a large difference
in the location of the He-flash ignition, but we suspect it
could be due to the choice of radiative and/or conductive opacities
adopted by~\citet{FIH90}, or differences in the treatment of
plasma-neutrino emission.
It should be emphasized that the value of $M(\epsilon^{\rm He}_{\rm
max})$ is very 
important for the occurrence of a H-flash.
The closer $M(\epsilon^{\rm He}_{\rm max})$ is to the border of the He
core, 
the higher is the probability for the
HECZ to reach the H-rich layer.

Soon after the He-flash sets in a convective shell appears, like in
the model computed by~\citet{FIH90}. However, in our
computation this convective zone reaches H-rich matter
$\approx10\,$yr after the He core flash, whereas in the work
by \citet{FIH90} the HEFM occurs immediately after the He-flash
($\approx10^{-3}\,$yr). 
This difference is due to the much
smaller distance between $M(\epsilon^{\rm He}_{\rm max})$ and the outer edge
of the H-exhausted region in \citeauthor{FIH90}'s \citeyearpar{FIH90} model 
compared to our computations. 

Fig.~\ref{fig1} shows the evolution of the energy release
produced by H and He burning during the major He core flash for two
models: the standard one and a model which does not experience a H-flash
(models S1 \& S3, see also Fig.~\ref{fig2}). 
At the onset of the H-flash, i.e.~when the H-burning luminosity exceeds 
that of He-burning, the outer border of the convective shell has
reached a mass shell with $X=0.055$. This value has to be compared
with the value  of $0.023$ listed by~\citet{FII00}, which actually belongs
to an $0.8\,M_\odot$ model, but which can be extrapolated safely to one
with $M=1\,M_\odot$.  

As soon as H is carried into the interior, it starts to be burnt at an
extremely high-rate due to the high temperatures in these regions, and
this deeply affects the following evolution of the star. This
behaviour is as in the computations by Fujimoto and coworkers.
We note that in this model experiencing a H-flash the amount of
H ingested during the HEFM phase is $6.5\times10^{-4}\,M_\odot$.

\begin{figure}
\center\includegraphics[width=\columnwidth]{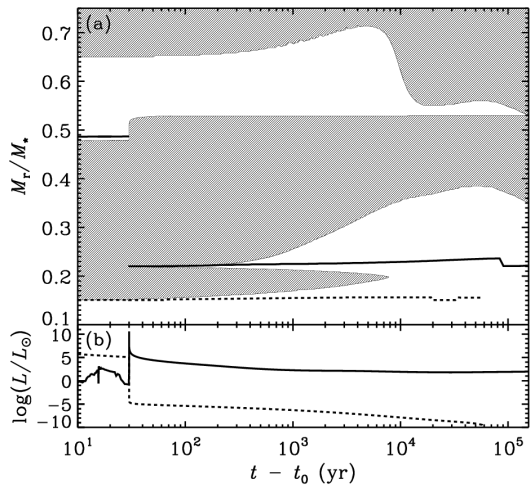}
\caption{(a) The development of the different convective zones --
indicated by shaded areas -- as a
function of time in our standard model. The solid and dashed lines
represent the mass shells of maximum energy release by H-
resp.\ He-burning. (b) The evolution of $L_{\rm H}$ (solid) and $L_{\rm
He}$ (dashed line) of the same model.\label{fig4}}
\end{figure}
The evolution of the convective zones
during the major He core flash is shown in Fig.~\ref{fig4}. 
The single HECZ separates into two zones
soon after the ingestion of H, because
two burning shells are developing 
\citep[an exhaustive description of this process can be found
in][]{HIF90}. 

This behaviour is in agreement with the computations performed by
\citet{HIF90},  
although the treatment of the driving mechanism, the CNO-burning,
is different: Hollowell and coworkers assumed that the abundances of
isotopes involved in the CNO-cycle are  locally in equilibrium,
and that all of the energy from the CN-chain is released whenever
a proton is captured on $^{12}$C, whereas we adopt a more accurate
treatment for the H-burning network (see \S~\ref{numtreat}).
In passing, we note that our models  support
the suggestion by~\citet{HIF90}, according to which a more realistic
treatment of the H burning should
have the only effect of producing a wider H-burning shell.

However, at odds with the results obtained by~\citet{HIF90}, in
our computations of the HEFM the He-burning luminosity
attains negligible values, and the structure is fully supported by the
H-burning shell. Besides, due to 
the H-flash the size of the He core is significantly reduced
in comparison with the value reached just before the He-flash.

About 100,000 years after the HEFM, the  convective envelope deepens
and  merges with the convective zone located above the H-burning
shell (the \lq{HCZ}\rq ), the lower boundary of which has, as
also shown by~\citet{HIF90}, moved outward. At the end of this
process, a huge amount of matter processed by both H and He burning
is carried to the surface of the star, polluting its initially
metal free chemical composition. Thus, the stellar surface is strongly
enriched by helium ($Y_{\rm surf}\approx0.5$) and the global metallicity
is equal to $Z=0.0131$. However, unlike a Pop.~I star, the most abundant
heavy elements at the stellar surface are the various isotopes of
carbon and nitrogen. The surface abundances of both elements
are listed in Table~\ref{tab1}.
It is evident that the He-flash induced H-flash, if operating in real 
Pop.~III stars, produces very significant changes in the chemical
composition of the star, preventing its identification as
a \lq{true}\rq\ metal-free star.

As a result of the dredge-up of metal-rich matter, the 
low-temperature opacity is increased drastically. This produces
an abrupt jump in the effective temperature visible in the
HR-diagram (Fig.~\ref{fig2}; indicated by the dotted part of the
track). 

The ceasing fusion of He after the H-flash yields
a new-born shell H-burning object, climbing up its own RGB.
However, this time the H-burning shell is moving through a 
helium-rich region and the energy production
rate is quite large in comparison with the previous evolution.
About 50 million years after the first He-flash, when
the He core mass is equal to $0.452\,M_\odot$, a second He-flash
develops at
$M_r=0.027\,M_\odot$. The smaller He core mass and the -- compared to
the first flash -- lower $M(\epsilon^{\rm He}_{\rm max})$
can both be explained by a substantial decrease of the central electron
degeneracy during the previous He ignition. This argument is 
supported by the strength of the second He-flash, which is
significantly weaker than that of the first one. 

In the subsequent evolution no further
mixing episode between the He-burning region and the outer envelope 
was found. The core He burning and later
phases will be described in \S~\ref{PostHe}. In the next section we
concentrate on the initial and physical conditions, which might
influence the development of the H-flash.

\subsection{The dependence of the H-flash on the initial
conditions.}\label{Hflash}

The results discussed in the previous section have clearly shown that,
at least for the ``standard'' model, our computations provide 
independent support for the results found earlier
by~\citet{HIF90} and by~\citet{FII00}. Nevertheless, there is the
important
difference between both investigations concerning
the location where the first He-flash starts.
As already stated in the previous section,
$M(\epsilon^{\rm He}_{\rm max})$  is 
closer to the border of the H-exhausted region in the models
provided by Fujimoto and coworkers than in ours.

Since the
position of maximum He-burning energy release is crucial in defining
whether the HEFM and the follow-up H-flash occur, we wish to
investigate if and to what extent the value of
$M(\epsilon^{\rm He}_{\rm max})$ depends on the initial conditions and on the
treatment of various physical processes.
Therefore, different models have been evolved through the major 
core He-flash. In detail, we have accounted for 
changes in the initial helium abundance, the stellar mass, the amount
of external pollution, the efficiency of atomic diffusion, and the
uncertainty in the mixing length theory, which is used for determining
the velocity of the convective cells in the adopted time-dependent
scheme. Furthermore, two different sets of equations of state and
opacities have been applied.
In Table~\ref{tab1}, the main properties of these models are listed.
The models which experience a H-flash are identified by $\dagger$.

{From} the data listed in this table, one notices that the
evolution through the HEFM phase strongly depends on
the initial conditions. For instance, increasing the
initial He abundance from 0.23 to 0.24  inhibits the
H-flash. The inward shift of $M(\epsilon^{\rm He}_{\rm max})$ is only
partially compensated by a decrease of the He core mass (Fig.~\ref{fig3}),
disfavouring a H-flash (see also Fig.~\ref{fig1}).

We have computed some sequences accounting also for atomic
diffusion as described by \cite{diffc} (models 'D'). These models
reveal that microscopic diffusion increases the probability for
a H-flash (compare models S3 and D3). Interestingly, by
artificially increasing the diffusion coefficients by a factor of 5,  
the internal structure of model D4, which
would not experience a H-flash,  
is modified in such a way that a H-flash does occur (model D+4).

Since the strength of the H burning, and probably also the
appearance of a H-flash, strongly depends on where inside the
HECZ the ingested protons are burnt --- the deeper
the burning occurs, the higher the burning rate --- we computed models
with different convective velocities and convective ``theories''.
In our standard model the velocity of the convective cells are
obtained from the mixing length theory \citep[MLT]{MWT}. We have
computed a model again, which usually experiences a H-flash, but decreasing
the MLT velocity by two orders of magnitude (model A1). This model
still shows a H-flash. In the same manner, increasing the MLT velocity
by two orders of magnitude in model S3, where no H-flash
occurs, does neither lead to the development of a H-burning runaway
(model A3). The convective velocity itself, therefore, is of no
influence on the flash.

Furthermore, we have  verified that our results do not depend on the 
convection theory applied. For this purpose, one sequence (C1)
has been computed using the convection theory by
Canuto-Mazzitelli \citep*[CM]{CM1,CM2}.  Again, a H-flash occurred
during the HEFM phase as in A1. Since the results are stable
against changes in the convection theory and against large variations
of the convective velocities, we consider them to be
reliable, despite the unavoidable approximations one has to make
when including time-dependent mixing in convective zones.

In view of the results discussed in Paper I, we have also checked
whether external pollution can affect the occurrence of a H-flash.
Therefore, we  computed additional models (P1 \& P3) by artificially
changing the chemical composition in the outer $0.01\,M_\odot$ of the
star, making it similar to the solar one (see Paper I for
more details).  We found
that  external pollution  disfavours the
development of a H-flash (see Table~\ref{tab1} and Fig.~\ref{fig3}).

All numerical experiments discussed so far have been performed by
using opacities where the metals are in scaled-solar proportions.
As described in \S~\ref{mainflash}, after the H-flash the ``metals''
in the convective zone above the H-burning shell (HCZ) consist almost entirely 
of carbon and nitrogen; therefore the boundaries
of the HCZ should be determined more accurately with opacity tables
that have  a more realistic heavy element distribution. It is
important to notice that the position of the lower  
boundary of the HCZ crucially determines
the amount of metals which are later dredged-up to the surface.

We performed some additional
experiments by using opacity tables taking into account the C-
and N-enhancement characteristic for these models. For this purpose
we computed opacity tables for the high-temperature
regime, considering various He abundances but including only
C and N in the heavy element distribution\footnote{These
opacity tables have been computed by using the tool available at
\url{http://www-phys.llnl.gov/Research/OPAL/index.html}, provided by
the Livermore Laboratory Opacity Group.}. 
In detail, we assumed that the C and N fractional abundances are both equal 
to 0.5 by mass within the metal group.
For lower temperatures the \citet{Alex} tables are used, which 
were only available for a weak carbon- and
nitrogen-enhancement ($3\times$solar, private
communication). Since the opacity evaluations
in the very outer regions are less crucial for the later mixing
process, we think that these tables are sufficiently accurate for
the present investigations.

We ran model O1 with these new opacity tables and
examined, in particular, if the boundaries of
the different convective zones change with respect to the models
with scaled-solar opacities.
No significant differences in the internal properties of the structure
and the final surface chemical composition could be detected.

A further test of the dependence of our results on the
adopted physical input consisted in computing a model (E1) using the
most accurate 
EOS presently available: the OPAL EOS \citep{OPEOS}. Since this
equation of state does not extend to the 
temperatures of He burning, it was extended by the simpler EOS
described in section~\ref{inphy}.  
The OPAL EOS is available for three metallicities ($Z=0.0,0.2,0.4$);
as for the ``standard'' opacity tables, metals are assumed 
to be scaled solar. We included this equation of state by
interpolating between the available tables at each stellar layer
to take into account the large variation of metal content in the
star. As a result, 
a stronger He-flash occurred in model E1 than in the standard case S1,
further increasing the probability of a H-flash during the HEFM.

In order to test the effect of mass loss, we changed the mass loss
rate (model B1 \& B2). 
No effect on the general behaviour of the HEFM process was found.
This is explained by recalling
that the outer envelope properties of RGB stars do not have much
influence on the internal properties.

As a general rule, summarizing the results shown in Fig.~\ref{fig3}
and listed in Table~\ref{tab1},  
{\em all those changes in the initial conditions and/or in the physical input, 
which can contribute to an increase of the electron degeneracy, and thus cause
the location of He ignition to move closer to the border of the He
core, favour the occurrence of a H-flash.}
This has the additional effect
that less energy released in the He-flash needs to be used for
expanding overlying layers. This statement is further
supported by the observation that the occurrence of a H-flash in
stars with smaller masses is less dependent on the initial conditions
and physical processes included. These stars intrinsically have a
more strongly degenerate electron gas in the center and thus the
development of a H-flash in these stars is less sensitive to the
various input parameters we investigated above. As an example, we
calculated model M1 and M3 with 0.82$\,M_\odot$, which both have a
larger maximum He-energy production ($L_{\rm He}^{\rm max}$) in the He
flash than almost all 1$\,M_\odot$ models and which both undergo a H
flash (see Table \ref{tab1}).


An important result of the experiments presented is 
that the final surface abundances of C and N in all models
experiencing a H-flash during the HEFM stage are quite
similar (they differ at most by a factor of 2,
cf.~Table~\ref{tab1}). This means that despite the difficulties in the
numerical treatment  
of the HEFM and the H-flash phases, the computations
provide quite robust results for
the relevant surface chemical abundances, which will be compared to
observations in \S~\ref{cobs}.

\section{The post He-flash evolution.}\label{PostHe}

\subsection{The core He-burning evolution.}

Irrespective of whether a H-flash occurs or not, as
soon as the model attains the thermal conditions necessary 
for the quiescent burning of He via the triple-alpha reaction, 
it displays the configuration typical of a Horizontal-Branch (HB) star.

Until now only one theoretical
investigation \citep{CCT} has addressed the problem of the evolution
of low-mass metal-deficient stars during the He-burning
phases. However, in that work the occurrence
of HEFM was neglected and the post He-flash evolution
was investigated without taking into account 
the changes in the stellar properties induced by this process.

In this section we analyze the evolution of a model which
has experienced a H-flash during the HEFM.
We consider a $1\,M_\odot$ model (S1) --- the
same one which we have defined as our standard model in the previous
discussions --- and the
$0.82\,M_\odot$ models B1 \& B2. The latter models have been
chosen because their age ($\approx 13$~Gyr) at the
Main Sequence Turn Off could be that of a contemporary Pop.~III star.

The evolution in the H-R diagram of the $1\,M_\odot$ model experiencing
the HEFM is shown in Fig.~\ref{fig2} (model S1). For comparison, the track during
the same evolutionary phase for model S3 (no H-flash) is
plotted (dashed line); filled circles indicate the
location of the Zero Age Horizontal Branch (ZAHB) for
both models. We notice that
despite the quite different evolution during the previous
core He-flash, no major difference in the
ZAHB luminosity ($\Delta\log(L/L_\odot)\approx0.1$) exists.
The ZAHB effective temperature is changed
only little, too ($\Delta\log{T_{\rm e}}\approx0.06$). The reasons why
the ZAHB location of the model experiencing a H-flash is slightly
brighter and cooler have been identified to be the larger He envelope
abundance and the larger surface metallicity, respectively. 
We also mention that the value of
the effective temperature for the star undergoing HEFM is probably not
precisely determined by our models, since the metal mixture of the
low temperature opacities does not exactly match the one at the
stellar surface. 

The time spent in core He burning 
is 77\,Myr for the ``standard'' model S1 and 51\,Myr for 
model S3 without HEFM. This difference is due to both the smaller He
core at the beginning of the central He-burning phase and the larger
efficiency of the H-burning shell in the model experiencing a H
flash. 

Interestingly, the evolutionary track in the H-R diagram of model S3,
which does not experience the H-flash, is quite similar to the results 
by~\citet{CCT}. However, the core He burning lifetime in their models
is longer by a factor of $\approx2$. This is mainly due to   
neglecting semiconvection during the quiescent He burning in our
computations. Ignoring 
this effect probably also caused the
occurrence of a blue loop which is not present in \citeauthor{CCT}'s
\citeyearpar{CCT} work. 

\begin{figure}[ht]
\includegraphics[width=\columnwidth]{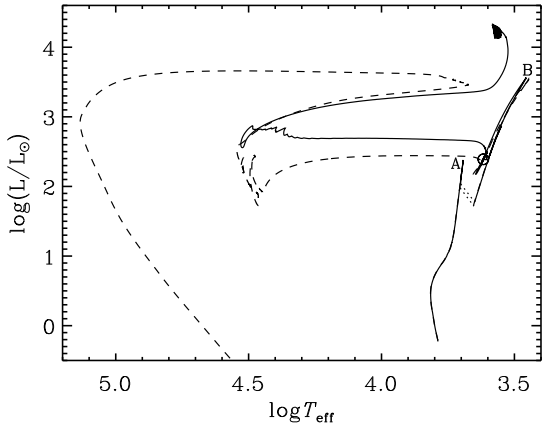}
\caption{The H-R diagram of $0.82\,M_\odot$ models experiencing
H-flashes, with Reimers' mass-loss ($\eta=0.4$, dashed
line) and without mass-loss (solid line). 
In Table~\ref{tab1} these models are denoted as B2
resp.~B1. The open circle indicates the --- almost
identical ---
ZAHB-positions of the two models.\label{fig5}}
\end{figure}

The evolution of the $0.82\,M_\odot$ model in the H-R diagram is shown
in Fig.~\ref{fig5}. In order to test the effect of mass-loss on the
overall evolution, the sequence
has been computed twice, first without mass-loss (model B2 in Table 1)
and then using a Reimers' mass-loss law with $\eta_{\rm Reimers}=0.4$ (model B1).
Due to the high mass loss in model B1
the mass of the star has been reduced to $0.63\,M_\odot$ at the
beginning of the HB evolution.
In spite of the large difference in their
total mass, the ZAHB location (open
circle in Fig.~\ref{fig5}) of both models is quite similar and
at $\log(L/L_\odot)\approx2.4$
and $\log{T_e}\approx3.6$. These values are a
consequence of the H-flash which increases the surface metallicity
($Z_{\rm surf}\approx 0.017$) in both models. Therefore, they behave as
metal-rich stars which are known to
show quite red ZAHB location for masses larger than
$\approx0.55\,M_\odot$ \citep[see for
instance][]{B97}. One notices that soon after the onset of
core He burning both models perform an excursion toward the blue
side of the H-R diagram. This behaviour can be explained by the quite
large helium abundances in their envelopes ($Y_{S}\approx0.5$).

The possibility that a metal-deficient stellar population produces a
significant number of RR-Lyrae pulsators has been investigated
by~\citet{CCT}. Neglecting the HEFM,
the ZAHB location of
metal-deficient stars is always hotter than the RR-Lyrae instability
strip, and one cannot expect to have any ZAHB pulsators. In addition, a
close inspection of the evolutionary paths in the H-R diagram revealed
that the possibility to form a significant number of variables from
the more evolved stars is negligible, too. In our scenario, accounting
for the HEFM process and a subsequent H-flash, once again the
possibility that metal-free stars
produce RR-Lyrae variables is quite negligible, but for the opposite
reason: metal-deficient HB stars experiencing the H-flash
have ZAHB locations too red in comparison with the RR-Lyrae instability
strip. In this respect, metal-deficient stars, the surface of which has been
polluted during the H-flash phase, behave
as ``true'' metal-rich stellar structures. The only possibility for obtaining
ZAHB pulsators is a quite strong, but improbable mass-loss rate
($\eta_{\rm Reimers}>0.4$), which  \lq{forces}\rq\ the ZAHB location to lie
within the instability strip. The post ZAHB evolution does not yield
an appreciable number of RR-Lyrae stars, because the
instability strip is crossed in a very short time
($\approx1-2$ Myr).


\subsection{The helium shell-burning phase.}

The high He abundance in the envelope of post H-flash stars
($Y\approx0.5$) leads to a large burning rate in the H shell. 
Therefore, at the end of core He burning the size of the H-rich
envelope is significantly reduced in comparison to models
which did not experience the H-flash at the RGB tip. Hence the
HB progeny of a formerly H-flashing star has a larger probability to
evolve as AGB-manqu\'e\ or post Early-AGB objects.

The evolution of low-mass metal-deficient stars during the Asymptotic
Giant Branch (AGB) has been investigated by~\citet{CCT} and
\citet{FII00}.
It was found that for these masses, a process
quite similar to the HEFM discussed previously can
occur at the beginning of the thermal-pulse phase. Hydrogen is mixed
into the convective zone driven by He-burning, causing -- like in the
H-flash phase at the tip of the RGB -- significant changes in the
surface chemical composition. 

\begin{figure}[b]
\includegraphics[width=\columnwidth]{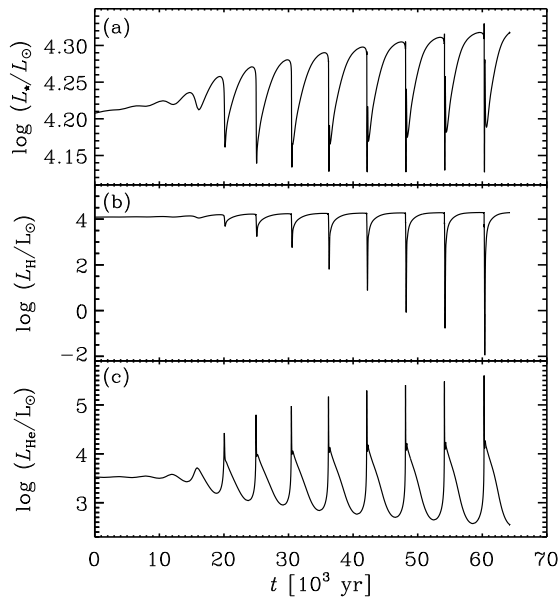}
\caption{The evolution of the total (a), the H-burning (b), and 
the He-burning luminosity (c) during the first 8
thermal pulses on the AGB for model B2 (see Table \ref{tab1}); this model
experienced a H-flash during  the HEFM phase.\label{fig6}}
\end{figure}

Because we have evolved model B2 (no mass-loss) through the AGB phase we
are now in the position to explore whether the H-flash at the RGB tip
affects the AGB evolution. In Fig.~\ref{fig6} we show the
evolution of the H-burning, the He-burning, and the total stellar luminosity
during the first 8 thermal pulses of model B2.
We have found an
interesting, but expected result: if a H-flash occurred at the
RGB tip, then the star behaves as a normal
thermal pulsing Pop.~I or II star with no further HEFM
episode on the AGB.

This behaviour is due to the dredge-up
of metals --- carbon and nitrogen --- into the stellar
envelope during the H-flash phase, with the consequence that 
during the following AGB
evolution the H-burning shell moves into a region no more
devoid of metals. 

\begin{figure}[ht]
\includegraphics[width=\columnwidth]{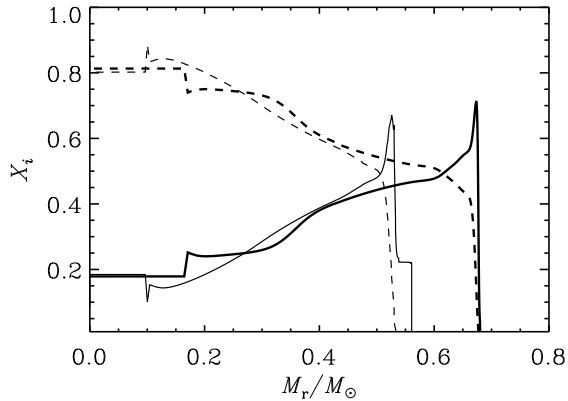}
\caption{The chemical stratifications of carbon (solid) and oxygen
(dashed line) within the
core during the 5th thermal pulse for $1\,M_\odot$
models with two different initial
chemical composition: metal-free (thick lines) and solar composition
(thin lines).\label{fig7}}
\end{figure}

Before concluding this section we briefly discuss
the carbon-oxygen (CO) abundance profile during the AGB phase. The CO
stratification at this stage corresponds closely 
to that inside the final white dwarf configuration, which 
is a key parameter to determine its cooling time
\citep[see, e.g.,][]{SDG97}.
In Fig.~\ref{fig7} we show the
carbon and oxygen profiles within the CO core at the 5th thermal pulse
for our standard model (S1) and for a
$1\,M_\odot$ star of solar composition.
In spite of the very different initial
metallicity, the CO profile is quite similar
throughout the core, and moreover it is
not affected by the occurrence of a H-flash during the RGB
evolution of the standard model.
A notable difference, however,
is the extension (in mass) of the core itself. The standard model shows a
much larger CO core, by about 0.1$\,M_{\odot}$. This should correspond to an
analogous difference for the final white dwarf masses, provided both
structures leave the AGB after approximately the same number of
pulses. 
We refrain from comparing in detail these profiles with corresponding results (for
solar metallicity stars) by \citet{SDG97}, because in our models we
not only neglect semiconvection during the quiescent He-burning phase,
but also use a different $^{12}{\rm C}(\alpha,\gamma)^{16}{\rm O}$
reaction rate. 

\begin{deluxetable}{cclllllr}
\tablecolumns{8}
\tablewidth{0pt}
\tablecaption{Observational properties for some selected theoretical models and
for two extremely metal-poor star.\label{tab2}}
\tablehead{
\colhead{Model} &  \colhead{Age [Gyr]} & \colhead{$T_{\rm eff}$ [K]} &
\colhead{$\log L/L_\odot$} & \colhead{[C/Fe]} & \colhead{[N/Fe]} &
\colhead{[Fe/H]} & \colhead{$^{12}$C/$^{13}$C}
}
\startdata
CS22892-052 & $15.6\pm4.6$~(1)     &   4760~(2)          &   2.5\tablenotemark{a}       &   1.1~(3)  &
1.0~(3)   &  -3.0~(3) & $\ga10$~(4)\\ 
CS22957-027 & \nodata     &   4840~(5)          &   1.7\tablenotemark{b}       &   2.2~(4)  &
2.0~(4)   &  -3.4~(4) & $10$~(4)\\ \tableline
B1    &  13.7  &   4500            &   1.71           &   3.9  &  4.1
&  -3.3 & 4.8\phm{~(4)} \\
      &  13.7  &   4700            &   2.4            &   4.0  &  4.2
&  -3.3 & 4.4\phm{~(4)} \\ \tableline
B2    &  13.7  &   4450            &   1.77           &   3.8  &  4.0
&  -3.2 & 4.8\phm{~(4)} \\
DP1   &  6.57      &   4570            &   1.68           &   2.2  &
2.5   &  -1.7 & 4.6\phm{~(4)} \\
\enddata
\tablenotetext{a}{Derived from $T_{\rm eff}$ and $\log g=1.3$ of Reference (2) by
assuming a 
mass of 0.82$\,M_\odot$.}  
\tablenotetext{b}{Derived like for CS22892-052 from $\log g=2.25$ of Reference (5)} 
\tablerefs{(1) \citet{CPK99}; (2) \citet{smcp:96}; (3) \citet*{nrb:97},
(4) \citet*{nrb97l}; (5) \citet*{BMB98}}
\end{deluxetable}

\section{Comparison with observations.}\label{cobs}

Our ultimate goal being to investigate the conjecture that the
observed ultra-metal poor stars are the first low-mass stars in our
Galaxy, we now confront our models with the observations. According to
\citet{beers:99} 10\% of the stars with $\mathrm{[Fe/H]} \le -2.5$ show a
carbon overabundance of $\mathrm{[C/Fe]} \ge 1.0$, this fraction
rising to 25\% at $\mathrm{[Fe/H]} \le -3.0$, with the excess
increasing to $\mathrm{[C/Fe]} = 2$ at the lowest metallicities
\citep{RBS99}. In addition, nitrogen is enhanced to similar
levels (where determined), while oxygen remains normal with respect to
iron, i.e.\ enhanced by $\approx 0.4$~dex as is typical for
Pop.~II. No standard stellar source is known for producing such
abundances \citep*{tww:95} and neither do so massive Pop.~III
supernovae (Heger \& Woosley 2000; private communication). We
therefore follow the idea that the peculiar C and N abundances stem from
H and He burning in the interior, while all heavier elements are the
result of pollution or accretion.  Our first prediction therefore would be that
only post He-flash objects can show enhanced C and N
abundances. Because the flash occurs at comparably low
luminosities in Pop.~III stars and the star experiences a rather
long-lived second RGB phase at lower $T_{\rm e}$ (due to the strongly
increased metal abundance in the envelope\footnote{calling such
post-flash stars {\em metal-poor} is appropriate if one refers to
true metals only}), such stars can also be found at rather moderate
RGB-brightness. 

As a first object to compare with we chose
CS22892-052, which is probably the best-observed star outside the
solar system. In
Table~\ref{tab2}, we list the stellar properties. The effective
temperature we took from \citep{smcp:96} as well as its gravity ($\log
g = 1.5$). Assuming that the mass is that of our model star with the
turn-off age being 13~Gyr ($0.82\,M_\odot$), we derive the luminosity
to be $\log L/L_\odot \approx 2.5$.  
Additionally, we add CS22957-027 \citep*{nrb97l}. 

We compare these observed
values with the ones obtained from our ``best'' models (B1 \& B2), whose
properties are listed in the next three rows of
Table~\ref{tab2}. 
We consider B1 and B2 to be the best theoretical counterparts of   
CS22892-052, since their mass is the one chosen to 
represent currently evolving Pop.~III stars (their age at the RGB tip
is 13.5~Gyr, consistent with actual estimates of the age of the
Universe, and with the value for CS22892-052 obtained by \citet{CPK99} from
nuclear chronology). 
From the B1-sequence we selected two models: the first one immediately after
the helium flash and the second one at a luminosity as estimated for
CS22892-052. The first one is also close to the luminosity of
CS22957-027.  These models are at the approximately correct locations in
the HRD. 
We have tuned the 
amount of polluted matter in order to match approximately the
observed [Fe/H] ratio in these stars. The predicted C and N abundance
ratios with respect to iron are listed in Table~\ref{tab2}, columns 5
and 6.

These theoretical predictions are 2--3 dex larger
than the observed values. Even if there are objects at that
metallicity that show enrichments in carbon of up to [C/Fe]=3, our
models still predict too much carbon and nitrogen. 
In fact, while the
observations indicate a spread in the overabundances, our models seem
to produce always the same amount of C and N, such that [C/Fe] and
[N/Fe] depend on the assumed pollution in  Fe, which does, however,
not influence the HEFM. Our models thus clearly fail to reproduce the
absolute amount of carbon and nitrogen in these stars and it remains to
be investigated 
whether an additional parameter controlling C and N
production can be found.

In Table~\ref{tab2} we list the same quantities also for a $1\,M_\odot$
model (DP1). It is evident that in this case the [C/Fe] and [N/Fe] ratios 
are in better agreement with observations, but this result
is just due to the fact that this model was polluted with more
material and therefore has a higher abundance of
iron at the
surface, which disagrees with the observed one. We add this model only
to underline the fact that the amount of polluting material hardly
influences the internal C and N production.

On the other side, it is reassuring that [C/N], which is 0.1 in
CS22892-052 and 0.2 in CS22957-027, is similar in our models ($\approx
-0.2$), and that effective temperature and luminosity can be matched
with models after the He-flash.

\section{Discussion and final remarks.}\label{dis}

In the present work we have investigated the evolution of low-mass
metal-free stars during the central ignition of He burning
at the RGB tip. We have used an improved
description of mixing and nuclear reaction network for these
peculiar stellar structures. This has allowed us to examine in
detail the occurrence and the properties of a mixing process induced
by the He-flash.

Our results provide independent support for the conclusions reached
by Fujimoto and coworkers. Actually, when the H-flash occurs, not only
the general properties of this phenomenon, but also its final consequences
on the surface chemical abundances, are quite similar to those
found by \citet{HIF90} and \citet{FII00}.
There is an important difference, however: whereas they claim
that the H-flash is a common property of all low-mass ($<1.0\,M_\odot$)
metal-deficient
stars, we have found that the occurrence of a H-flash depends
on the adopted initial conditions, like
mass and He abundance, as well as on the assumptions made for the
treatment of some physical process like diffusion and external
pollution. Our computations disclose that the occurrence of a H-flash
is mainly governed by one parameter: the position of the maximum  
He-burning energy release in the He-flash.
The deeper inside the star this point lies, the lower is
the probability that the convective zone developing at the He-flash
ignition can penetrate into the hydrogen-rich envelope, thereby carrying
down protons and triggering a H-burning runway.

A first comparison with CS22892-052 and CS22957-027 revealed a good
match of effective 
temperature and luminosity and the relative abundances of carbon and
nitrogen, but a severe overproduction of these elements, which is
independent of the assumed amount and composition of the polluting
material. 
Even if the H-flash phenomenon provides an
interesting working scenario for interpreting the observed
abundance patterns in extremely metal-deficient stars, it is not yet
able to provide a detailed match to the observations.
There are several possible interpretations of this
failure, and we are now going to comment on some of them.

In the H-flash N is produced in the CN-chain by the fusion of carbon
with hydrogen. Thus, any process destroying part of the carbon
before the HEFM occurs also reduces the final nitrogen abundance.
Obviously, with the rate of the $^{12}{\rm
C}(\alpha,\gamma)^{16}{\rm O}$ reaction being affected by a large
uncertainty \citep{B96}, a possible process to reduce the carbon
abundance can be identified. 
However, our computations show that the mean
temperature in the He-burning region is far too low for
the $^{12}{\rm C}(\alpha,\gamma)^{16}{\rm O}$ reaction being
efficient. Hence the demanded changes in the $^{12}{\rm
C}(\alpha,\gamma)^{16}{\rm O}$-rate to process considerable amounts of
carbon, would be much larger than the experimental uncertainties.
In addition, the surface oxygen abundance would be increased
considerably, in contrast to the observations, which reveal an oxygen
content typical for Pop.~II stars.

Another possibility for reducing the amount of C and N dredged up to
the surface is an outward shift of the lower boundary of the
convective zone driven by the H-flash, when merging with the
convective envelope.  It is however unclear how this could be done,
because in all our models, even in the one containing the opacity with
C-and N-enhanced mixture --- which we consider to be the most reliable
ones --- the convective boundaries are located almost at the same
position.

The third, and in our opinion the most promising way out of the high C
and N abundances lies in an improvement of the crude one-dimensional
description used 
to calculate the violent H burning in the HECZ zone. Possibly, we
overestimate the convective velocity, and thus the penetration depth of
hydrogen. Model A1 with reduced velocities of the convective elements
shows the tendency of smaller carbon and nitrogen abundances
(about 30\% less than in the standard model S1, cf.~Table~\ref{tab1}).
In addition, the $^{12}{\rm C}/^{13}{\rm C}$-ratio, which in all
H-flashing models is about $4.5\pm0.5$, is 5.6 in model A1,
which again points toward the experimental value of $\ga10$. 
A similar result might be obtained if the
penetration of the HECZ into the H-rich layers could be delayed until the
temperature in the He-burning shell has diminished.
We add that in 
the first two possibilities presented above the 
ingested hydrogen would be burnt under the same conditions yielding
the same lower values for $^{12}{\rm C}/^{13}{\rm C}$.

Nevertheless, since the dredge-up of C and N to the surface is a
consequence of the huge amount of energy released by hydrogen fusion,
and thus by the production of nitrogen, a minimum value for the
nitrogen abundance must exist. Below this value, the H-burning energy
release is not sufficient to change the subsequent evolution of the
star and no dredge-up would occur. Further
investigations are required for deciding whether the observed N and C
abundances can be reproduced by a modified mixing/burning description
or by multidimensional simulations.

We point out a further potential problem that might
become evident with larger samples of well-investigated metal-poor
stars: since we predict carbon-enhancement due to internal production
only during the He-flash, stars on the main sequence or near the
turn-off should not display the same abundance pattern. In
addition, above the post-flash luminosity ($\log L/L_\odot
\approx 1.5$) the number of post-flash to pre-flash stars should
reflect the evolutionary timescales (since the mass is approximately
constant, assuming similar ages for all such halo stars); the ratio
should be around 1:5 rising with increasing luminosity. This number is
not too far from the observed numbers, but it is unknown to what extent
observational biases might have influenced this. It is therefore extremely
important to get larger samples with additional information about the
stellar brightness. 

Although our models face severe problems when comparing them to the
best-observed ultra metal-poor star, CS22892-052, they still offer the
only source for producing the observed C and N abundance patterns. Due
to the mixing of protons into He-burning regions, $s$-process elements
might result from the HEFM. We will investigate this and the evolution
of light elements ($^7$Li) in future work.

\acknowledgements{This work was supported by a DAAD/VIGONI grant.
One of us (S.C.) warmly thanks for the financial support by
MURST (Cofin2000) under the scientific project: ``Stellar observables
of cosmological relevance''.
A.W. is grateful for the hospitality at the Institute for Advanced Study 
and Princeton Observatory and for a Fulbright fellowship which allowed
visiting both places. Helpful discussions with T.~Beers, C.~Sneden, 
J.~Cowan and J.~Truran and the very careful reading of the manuscript
by H.~Ritter are acknowledged.}

\newcommand{\singlet}[1]{#1}

\end{document}